\documentclass{article} 
\usepackage{iclr2023_conference_tinypaper,times}

\usepackage{amsmath,amsfonts,bm}

\def\eqref#1{equation~\ref{#1}}

\def\1{\bm{1}}

\DeclareMathAlphabet{\mathsfit}{\encodingdefault}{\sfdefault}{m}{sl}
\SetMathAlphabet{\mathsfit}{bold}{\encodingdefault}{\sfdefault}{bx}{n}

\usepackage{hyperref}
\usepackage{url}
\usepackage{booktabs}
\usepackage{graphicx}

\title{Learning Disentangled Audio Representations through Controlled Synthesis}

\author{Yusuf Brima, Ulf Krumnack,  Simone Pika \& Gunther Heidemann\\
Institute of Cognitive Science\\
Universität Osnabrück\\
Wachsbleiche 27, D-49090 Osnabrück, Germany\\
\texttt{\{ybrima,krumnack,spika,gheidema\}@uos.de} \\
}
\iclrfinalcopy 
\begin{document}
\maketitle
\begin{abstract}
This paper tackles the scarcity of benchmarking data in disentangled auditory representation learning. We introduce \textit{SynTone}, a synthetic dataset with explicit ground truth explanatory factors for evaluating disentanglement techniques. Benchmarking state-of-the-art methods on SynTone highlights its utility for method evaluation. Our results underscore strengths and limitations in audio disentanglement, motivating future research.
\end{abstract}

\section{Introduction}
Disentangled representation learning shows promise for creating useful data encodings. However, it remains underexplored for auditory data, despite potential benefits~\citep{locatello2019challenging,higgins2017beta,kim2018disentangling}. Disentangled representations refer to learned encodings (\textit{codes} for short) where explanatory factors are encoded in \textit{independent}, \textit{compact}, and \textit{informative} dimensions. These \textit{factorized} representations offer an improved generalization, explainability, reduced sample complexity, and avoidance of shortcut learning~\citep{carbonneau2020measuring}.

Progress in auditory disentanglement is hampered by the lack of suitable benchmarking datasets for evaluation~\citep{hsu2017unsupervised,mo2019semantic}. Auditory signals present unique challenges due to their \textit{high dimensionality}, \textit{temporal dynamics}, and \textit{rich hierarchical} structure. To help address this evaluation gap, we introduce \textit{SynTone}: a basic dataset of synthetic tones curated for quantitatively assessing disentangled representation learning techniques in the acoustic domain.
We used this new dataset to conduct a benchmarking study comparing state-of-the-art (SOTA) disentanglement frameworks. Performance was measured using \textit{supervised disentanglement metrics}. Our goals were to (i) demonstrate SynTone's utility for method evaluation/comparison, and (ii) highlight the strengths and limitations of audio disentanglement algorithms to help guide future research.
\section{Methodology}

\subsection{SynTone Dataset}
Following ideas from disentanglement in the vision domain~\citep{dsprites17,3dshapes18}, we develop SynTone - a simple synthetic audio dataset with explicit ground truth factors to facilitate disentanglement analysis. SynTone comprises $32,000$ 1-second audio samples (16kHz sampling rate) synthesized by systematically varying three generative parameters: timbre $\mathbb{T}$ (sine, triangle, square, sawtooth waveforms), amplitude $\mathbb{A}$ (20 discrete levels from 0 to 1), and frequency $\mathbb{F}$ (400 discrete steps from $440$ to $8000$Hz). The dataset structure is formally: $\mathbb{T} \times \mathbb{A} \times \mathbb{F}$, with each sample having a unique $(T \in \mathbb{T}, A \in \mathbb{A}, F \in \mathbb{F})$ tuple defining its generative factors. Having full knowledge of the data generative process allows quantitative assessment of models' effectiveness at learning the ground truth factors from the audio samples. This dataset aims to provide a benchmark for the analysis of disentanglement techniques on a perceptual signal domain where the factorized generative sources are specified.

\subsection{Modeling}
We employ four VAE-based models to disentangle SynTone's ground truth factors: vanilla VAE~\citep{kingma2013auto}, $\beta$-VAE~\citep{higgins2017beta}, Factor-VAE~\citep{kim2018disentangling}, and $\beta$-TCVAE~\citep{chen2018isolating}. The use of VAEs offers interpretable latent spaces, and these model variants introduce objectives specifically targeting greater independence among latent factors. SynTone's rich ground truth provides a basis for quantifying the success of recovering explicitly defined variations. Our focus lies on testing the explanatory disentanglement capabilities afforded by VAE techniques, in contrast to assessing the sample quality of less interpretable GANs~\citep{chen2016infogan}.

To evaluate the disentanglement achieved by these different approaches, we apply various metrics that quantify each model's ability to recover ground truth factors, emphasizing \textit{different desired properties} of disentanglement. These properties include \textit{modularity}, where variation in one factor does not affect other factors (i.e., there is no causal effect between them); \textit{compactness}, where a factor should be explained by a minimal subset of the latent space, ideally a single dimension; and \textit{explicitness}, where learned codes should be semantically meaningful. Specifically, we use the Mutual Information Gap (MIG)~\citep{chen2018isolating} and Attribute Predictability Score (SAP)~\citep{kumar2017variational} to measure compactness. We also use Joint Entropy Minus Mutual Information Gap (JEMMIG)~\citep{do2019theory} and Disentanglement, Completeness, and Informativeness MIG (DCIMIG)~\citep{sepliarskaia2019not} to assess \textit{holisticness}, as they capture modularity, compactness, and explicitness properties in a single score. Moreover, the code modularity is measured with the Modularity score~\citep{ridgeway2018learningfstatistic}. For a more in-depth analysis of disentanglement metrics, refer to~\citet{carbonneau2020measuring}.
\section{Experiments}
We trained four model architectures on SynTone: vanilla VAE, $\beta$-VAE, Factor-VAE, and $\beta$-TC-VAE. To facilitate training, the audio data is transformed into a time-frequency domain, resulting in a 2D input for our networks (refer to appendix~\ref{input_preprocess}). All variational autoencoders used the same convolutional architecture (detailed in appendix~\ref{model_architecture}), with variations only in the objective function. Extensive exploration of hyperparameter settings for the different models was conducted, and only the best-performing configuration was adopted for the final experiment. Table~\ref{tab:disentanglement-metrics} provides a summary of these metrics for each model.
\begin{table}[t!]
    \centering
    \begin{tabular}{lcccccc}
        \toprule
        \multicolumn{1}{c}{} & \multicolumn{2}{c}{\underline{Compactness}} & \multicolumn{2}{c}{\underline{Holistic}} & \multicolumn{1}{c}{\underline{Modularity}} \\
        Model & MIG & SAP & DCIMIG & JEMMIG & Mod.~Score \\
        \midrule
        Vanilla VAE & \textbf{0.280 $\pm$ 0.0} & \textbf{0.045 $\pm$ 0.000} & \textbf{0.005 $\pm$ 0.000} & 0.300 $\pm$ 0.0 & 0.651 $\pm$ 0.031 \\
        $\beta$-VAE & 0.184 $\pm$ 0.0 & 0.035 $\pm$ 0.000 & 0.004 $\pm$ 0.000 & 0.201 $\pm$ 0.0 & 0.750 $\pm$ 0.022 \\
        Factor-VAE &  0.118 $\pm$ 0.0 & 0.035 $\pm$ 0.000 & 0.003 $\pm$ 0.000 & \textbf{0.126 $\pm$ 0.0} &\textbf{ 0.762 $\pm$ 0.014} \\
        $\beta$-TCVAE & 0.166 $\pm$ 0.0 & 0.023 $\pm$ 0.000 &\textbf{ 0.005 $\pm$ 0.000} & 0.183 $\pm$ 0.0 & 0.747 $\pm$ 0.016 \\
        \bottomrule
    \end{tabular}
    \caption{Disentanglement metrics with standard deviation for different models over 10 evaluation runs each. Values are presented as mean $\pm$ standard deviation.}
    \label{tab:disentanglement-metrics}
\end{table}


In our analysis, the vanilla VAE's superior performance in compactness metrics (MIG, SAP) and comparable DCIMIG to $\beta$-TCVAE was unexpected, especially as compactness is of lesser practical importance. Its simpler structure may better isolate factors in the SynTone dataset. Factor-VAE led in JEMMIG and Modularity, as expected from its training approach. The underperformance of $\beta$-VAE and $\beta$-TCVAE in certain metrics suggests their loss functions might not be ideally suited for SynTone, emphasizing the need for dataset-specific approaches in audio-disentangled representation learning. For deeper analysis, see the qualitative results in the appendix~\ref{reconstruction_and_sampling}.
\section{Conclusion}
Our work addresses the unexplored realm of disentangled representation learning in audio by introducing SynTone, a dataset with explicit ground truth factors. We evaluated state-of-the-art methods, revealing the challenges in achieving comprehensive disentanglement in audio representations. While the vanilla VAE demonstrates robust compactness, the Factor-VAE excels in modularity. However, limitations persist, emphasizing the need for further research in audio disentanglement. Our dataset proves valuable for method evaluation, though potential extensions to more diverse datasets and real-world scenarios warrant exploration. Our findings provide insights for future efforts in enhancing disentangled audio representation learning.
\subsubsection*{Acknowledgements}
This work was funds of the research training group (RTG) in ``Computational Cognition'' (GRK2340) provided by the Deutsche Forschungsgemeinschaft (DFG), Germany, and an EU-Consolidator grant (772000, TurnTaking).

\subsubsection*{URM Statement}
The authors acknowledge that at least one key author of this work meets the URM criteria of the ICLR 2024 Tiny Papers Track.


\bibliography{iclr2023_conference_tinypaper}
\bibliographystyle{iclr2023_conference_tinypaper}
\appendix
\section{Appendix}
\subsection{Model Input Preprocessing}
\label{input_preprocess}
The input representation to each model is a mel-spectrogram transformation with a window size of $2024$ and a hop size of $512$, aligning with speech processing best practices. Mel-spectrograms efficiently encapsulate audio details into perceptually relevant frequency bands. The encoder outputs parameterize a multivariate Gaussian distribution over latent variables, guided by a $\beta$ hyperparameter during training for incentivizing disentangled dimensions. Supervised by $\mathbb{T}$, $\mathbb{A}$, and $\mathbb{F}$ generative factors, designated latent dimensions are manipulated post-convergence to control amplitude, timbre, and frequency in generated samples. Performance assessment includes quantitative and qualitative evaluations of disentanglement metrics, reconstructions, and sample quality respectively.
\subsection{Model Architecture}
\label{model_architecture}
Our VAE model applies four convolutional layers to extract hierarchical spatial features from input mel-spectrograms, followed by two fully connected layers to parameterize a Gaussian latent code capturing explanatory factors of variation. The decoder mirrors this pathway using transpose convolutions to recover the original input dimensions from sampled codes. All layers use leaky ReLU activations (slope=0.01) for faster convergence, with no batch norm for clearer disentanglement analysis. We strategically upsample in the decoder to balance reconstruction quality without overparameterization. This convolutional VAE structure provides an interpretable latent space to analyze captured generative factors while modeling complex mel-spectrogram dependencies. By parameterizing the distributions, we additionally obtain useful uncertainty information.

Building upon this, the full implementation details, including the architecture, training procedures, and hyperparameters used in our VAE models, are meticulously documented. This comprehensive resource is designed to assist in replicating or building upon our work. The provided link offers access to our codebase \href{https://anonymous.4open.science/r/SynTone/}{VAE Models Full Implementation}. 
\subsection{Reconstruction and Sampling}
\label{reconstruction_and_sampling}
We visually assess models' performance through reconstructions of held-out mel-spectrograms and novel samples synthesized by decoding random codes. Reconstructions indicate how effectively the data compression is captured. Sampling assesses generative modeling abilities when guided solely by the learned latent distributions. Comparing the similarity of reconstructions and coherence of samples provides intuitive insight into encoding quality between models, complementing metric-based evaluation.
\begin{figure}[!h]
    \centering
    \includegraphics[width=1.2\textwidth]{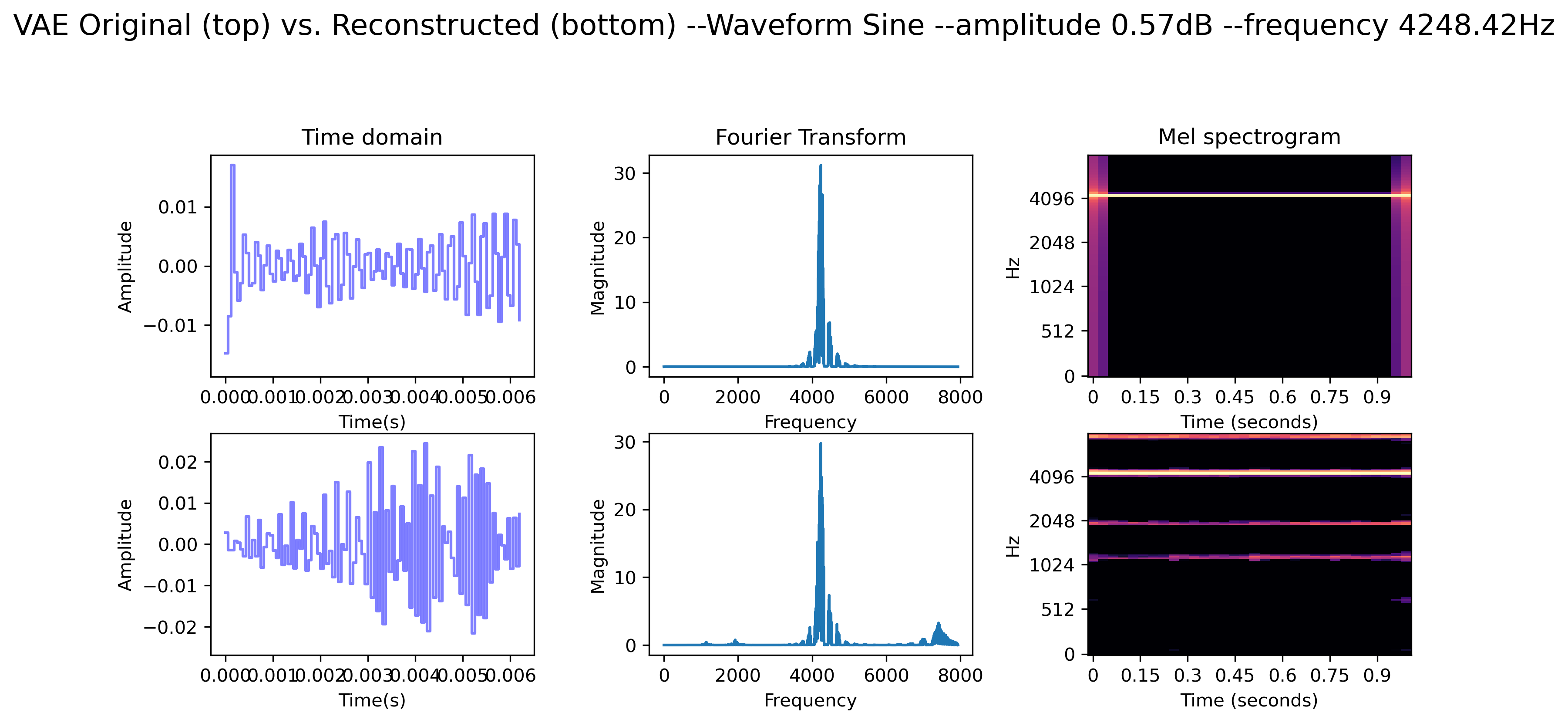}
    \caption{Time-domain representation (left), Fourier transforms (middle), and Time-Frequency (Mel-spectrogram) representation (right) for original and reconstructed samples using the VAE model. This model shows a lower reconstruction quality as indicated in all plots due to the addition of noise frequencies around 7000Hz as well as lower harmonic partials as portrayed in the mel-spectrogram.}
    \label{fig:Original_vs_Reconstructed_VAE}
\end{figure}


\begin{figure}[!h]
    \centering
    \includegraphics[width=\textwidth]{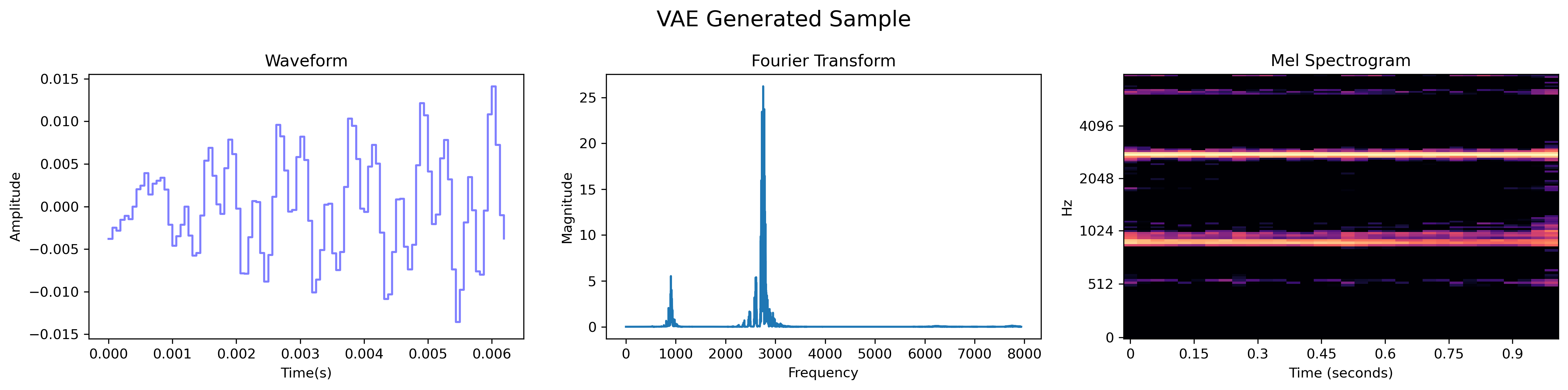}
    \caption{Visualization of generated audio samples using the VAE model. The left column depicts the time-domain representation, the middle column illustrates the Fourier transforms, and the right column showcases the Time-Frequency (Mel-spectrogram) representation. This plot shows that the model can isolate is single major frequency band close to 3000Hz as illustrated by the Fourier domain periodogram.}
    \label{fig:Generated_sample_VAE}
\end{figure}


\begin{figure}[!h]
    \centering
    \includegraphics[width=1.2\textwidth]{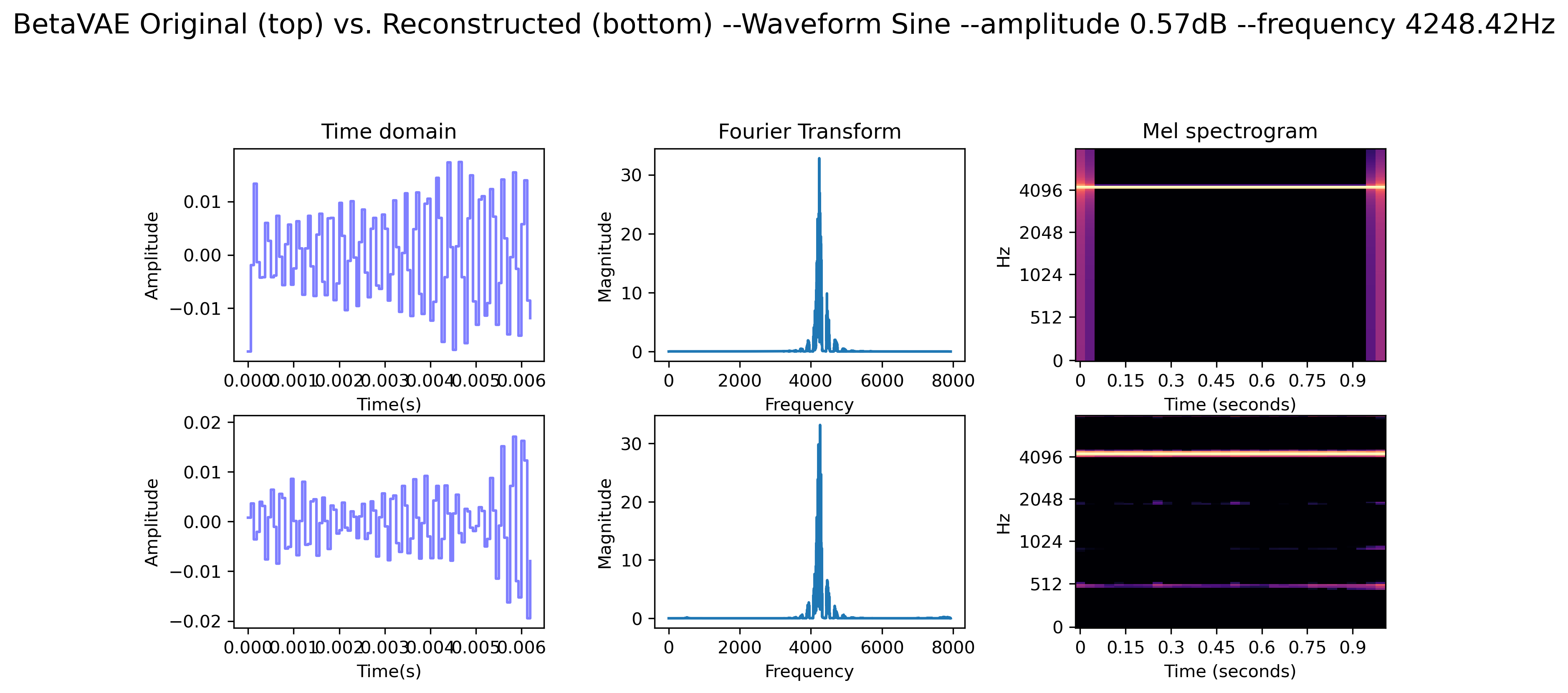}
    \caption{Time-domain representation (left), Fourier transforms (middle), and Time-Frequency (Mel-spectrogram) representation (right) for original and reconstructed samples using the $\beta$-VAE model. This model achieved a suitable reconstruction quality compared to the vanilla VAE as shown in both the Fourier space where the frequency factor of 4248.42Hz is correctly isolated.}
    \label{fig:Original_vs_Reconstructed_BetaVAE}
\end{figure}
\begin{figure}[!h]
    \centering
    \includegraphics[width=\textwidth]{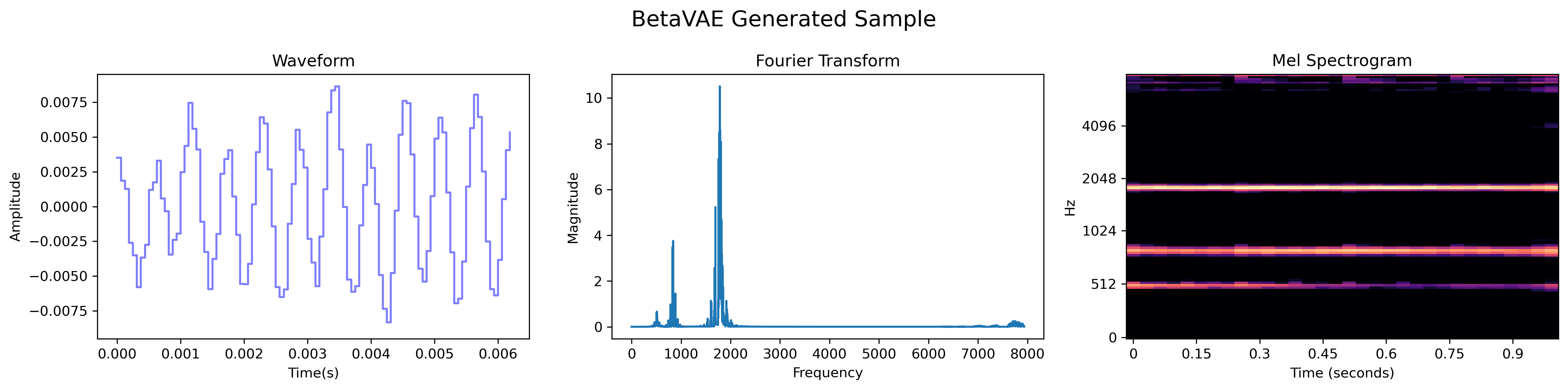}
    \caption{Visualization of generated audio samples using the $\beta$-VAE model. The left column depicts the time-domain representation, the middle column illustrates the Fourier transforms, and the right column showcases the Time-Frequency (Mel-spectrogram) representation. This figure, however, shows minor harmonic partials close to 1000Hz while the major frequency component is also predominant in both the periodogram and mel-spectrogram plots.}
    \label{fig:Generated_sample_BetaVAE}
\end{figure}

\begin{figure}[!h]
    \centering
    \includegraphics[width=1.2\textwidth]{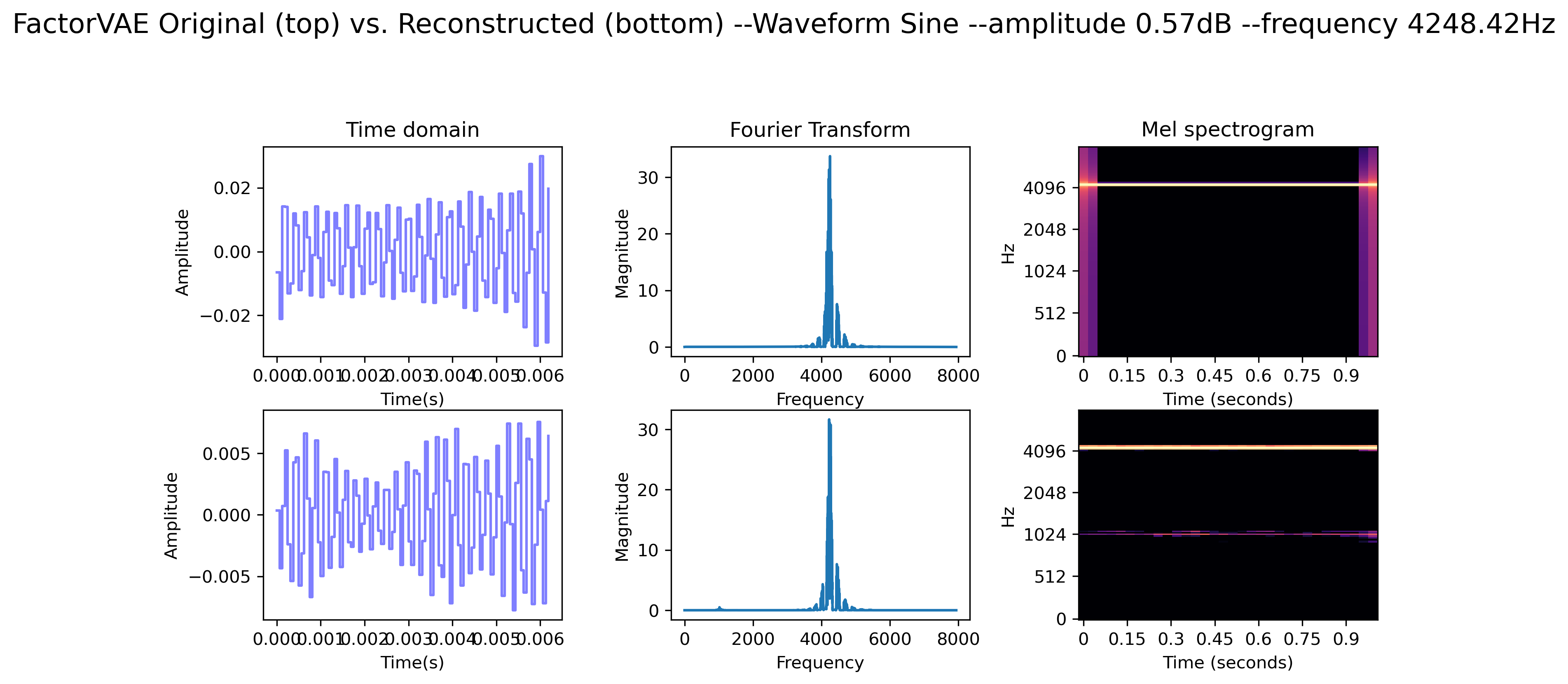}
    \caption{Time-domain representation (left), Fourier transform (middle), and Time-Frequency (Mel-spectrogram) representation (right) for original and reconstructed samples using the FactorVAE model. This model achieved the best reconstruction quality for an assessment of audio representation in both the Fourier space where the frequency factor of 4248.42Hz is correctly isolated.}
    \label{fig:Original_vs_Reconstructed_FactorVAE}
\end{figure}
\begin{figure}[!h]
    \centering
    \includegraphics[width=\textwidth]{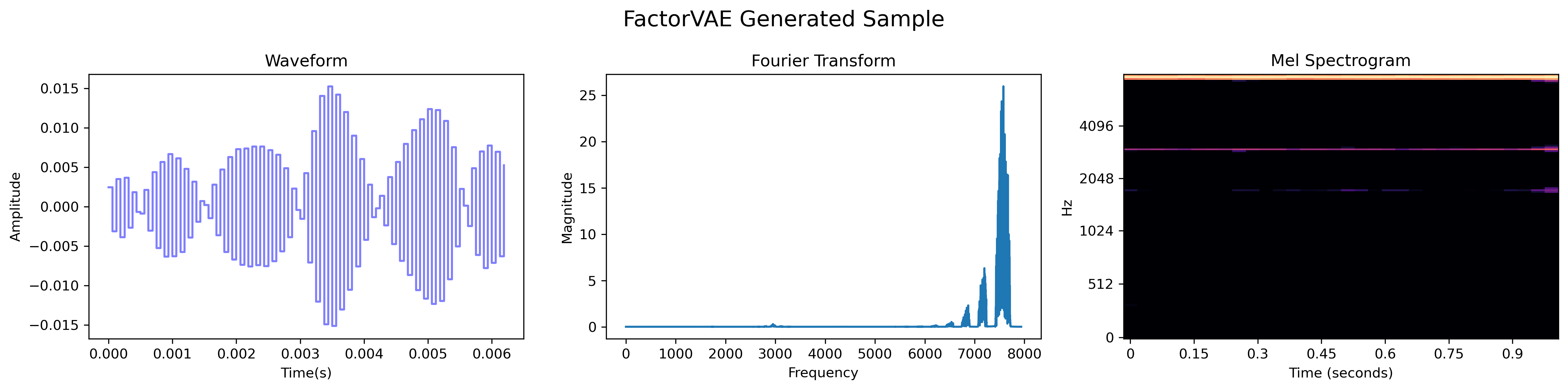}
    \caption{Visualization of generated audio samples using the Factor-VAE model. The left column depicts the time-domain representation, the middle column illustrates the Fourier transforms, and the right column showcases the Time-Frequency (Mel-spectrogram) representation. The model can decode a complex latent code that is smoothly represented in both the Fourier plot and mel-spectrograms with minimal noise levels.}
    \label{fig:Generated_sample_FactorVAE}
\end{figure}

\begin{figure}[!h]
    \centering
    \includegraphics[width=1.2\textwidth]{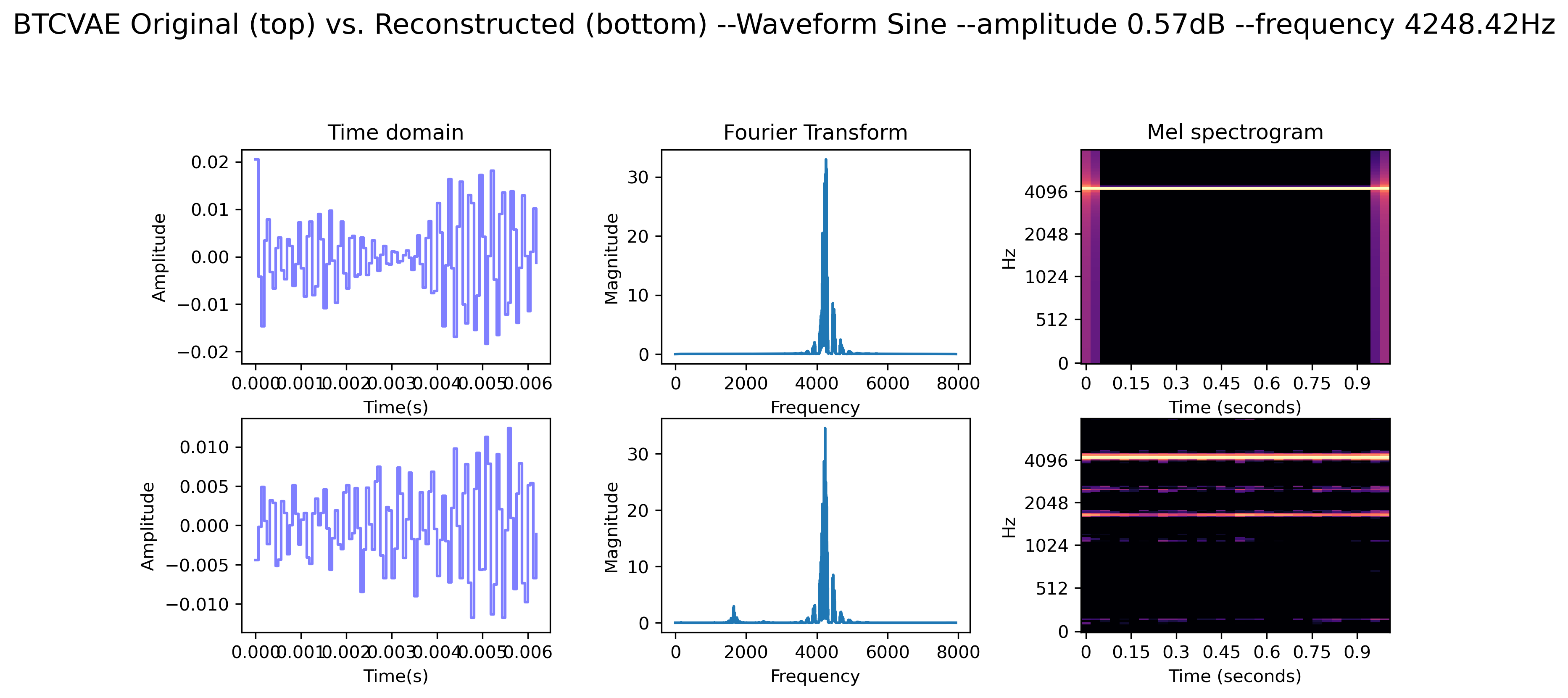}
    \caption{Time-domain representation (left), Fourier transforms (middle), and Time-Frequency (Mel-spectrogram) representation (right) for original and reconstructed samples using the $\beta$-TCVAE model. Like the vanilla VAE, this model also suffers from low-frequency noise in encoding and decoding the audio signal.}
    \label{fig:Original_vs_Reconstructed_BTCVAE}
\end{figure}

\begin{figure}[!h]
    \centering
    \includegraphics[width=\textwidth]{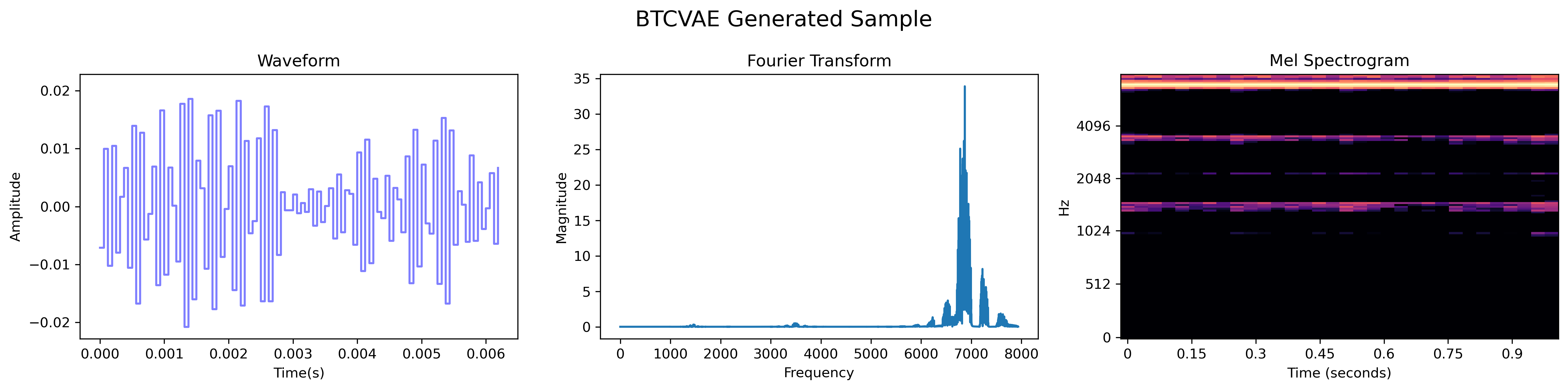}
    \caption{Visualization of generated audio samples using the $\beta$-TCVAE model. The left column depicts the time-domain representation, the middle column illustrates the Fourier transforms, and the right column showcases the Time-Frequency (Mel-spectrogram) representation. Like the Factor-VAE decoder, this model is also able to decode complex waveforms however, it has more predominate noise levels as indicated in the plots where minor frequency contributions are visible.}
    \label{fig:Generated_sample_BTCVAE}
\end{figure}

\begin{figure}[!h]
    \centering
    \includegraphics[width=\textwidth]{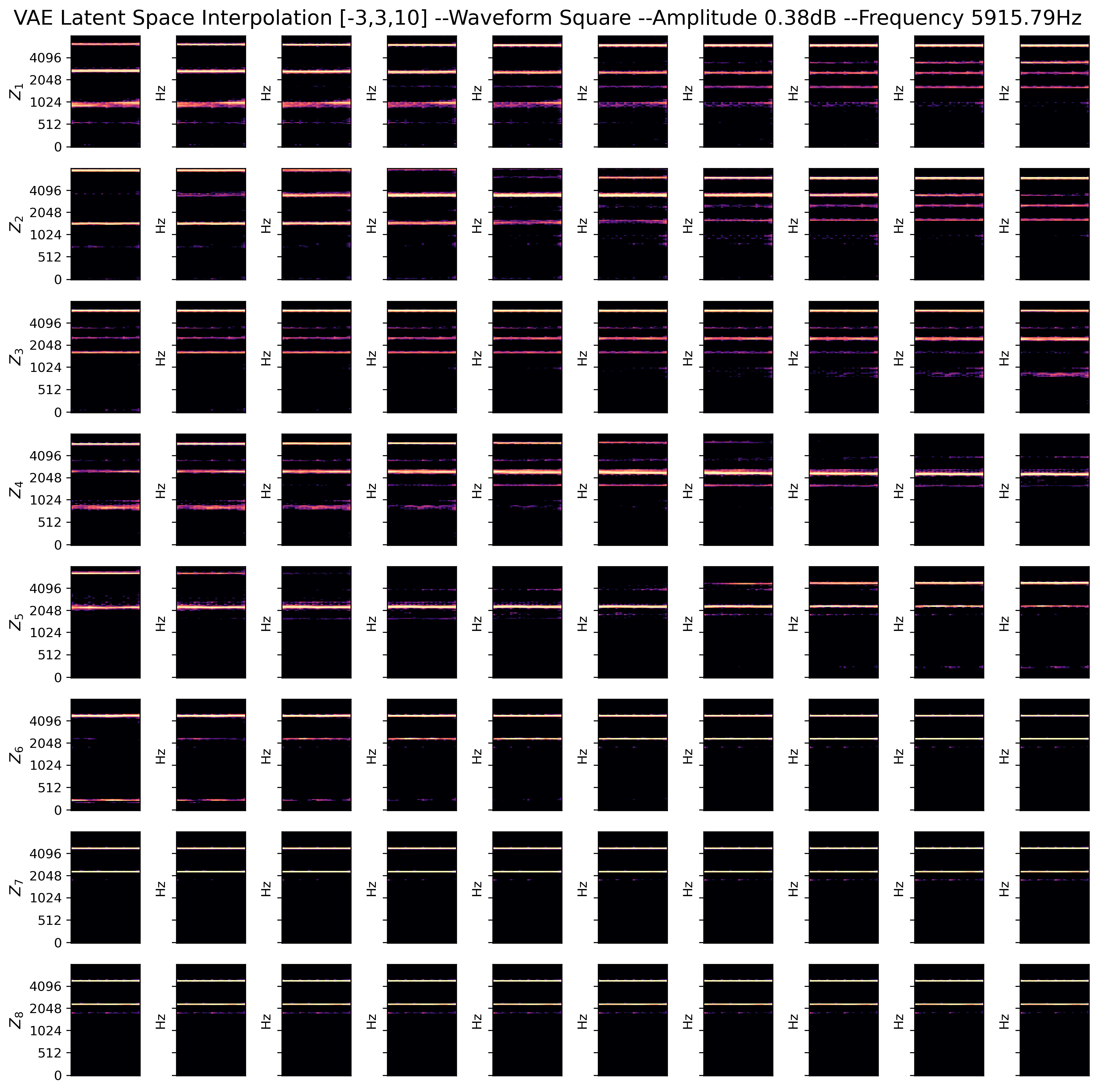}
    \caption{Latent space interpolation for the vanilla VAE model, showcasing variations in a single latent code while others are fixed. Each row represents 10 discrete samples per latent code.}
    \label{fig:VAE_Latent_Space_Interpolation}
\end{figure}

\begin{figure}[!h]
    \centering
    \includegraphics[width=\textwidth]{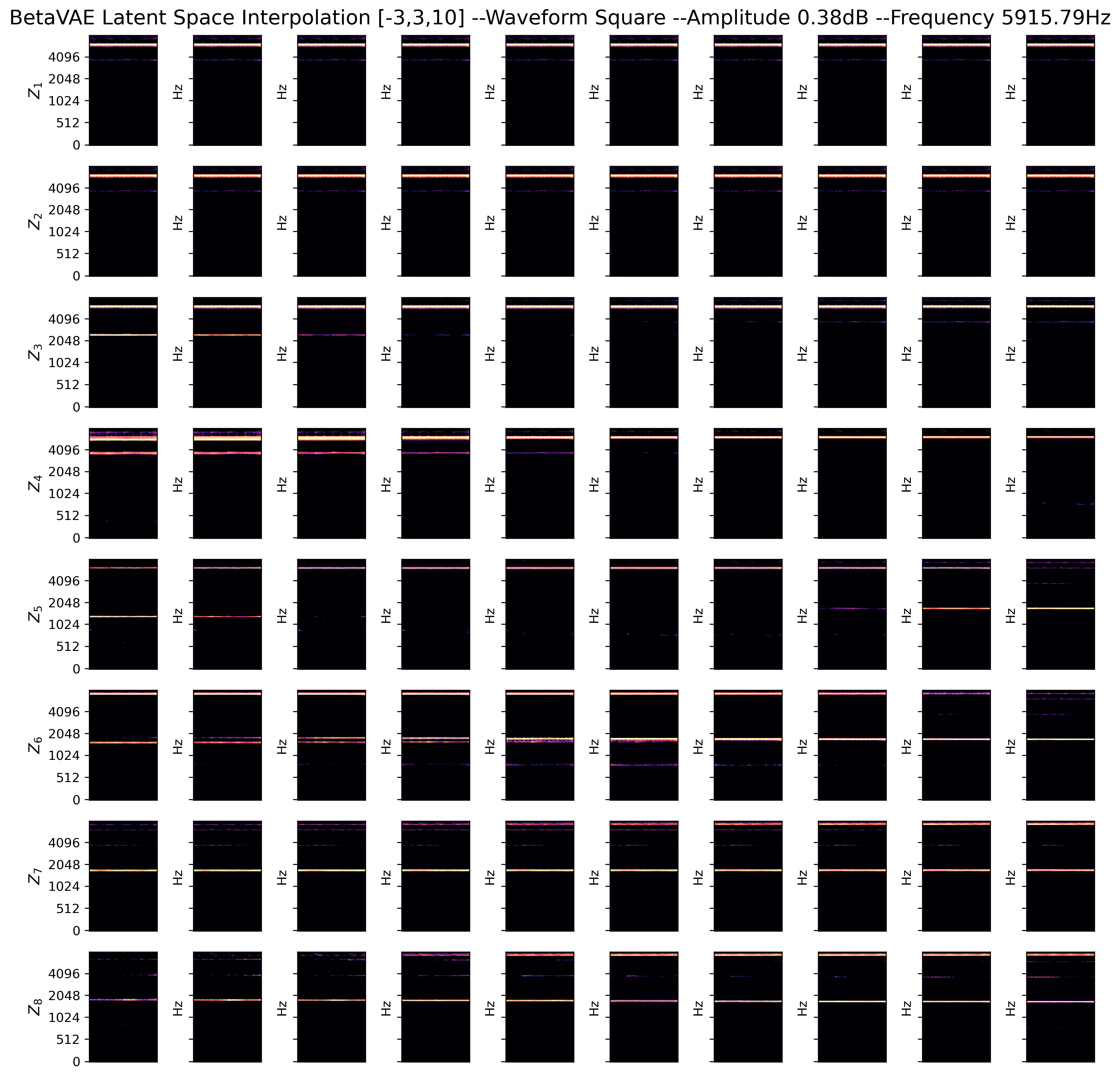}
    \caption{Latent space interpolation for the $\beta$-VAE model, showcasing variations in a single latent code while others are fixed. Each row represents 10 discrete samples per latent code.}
    \label{fig:BetaVAE_Latent_Space_Interpolation}
\end{figure}

\begin{figure}[!h]
    \centering
    \includegraphics[width=\textwidth]{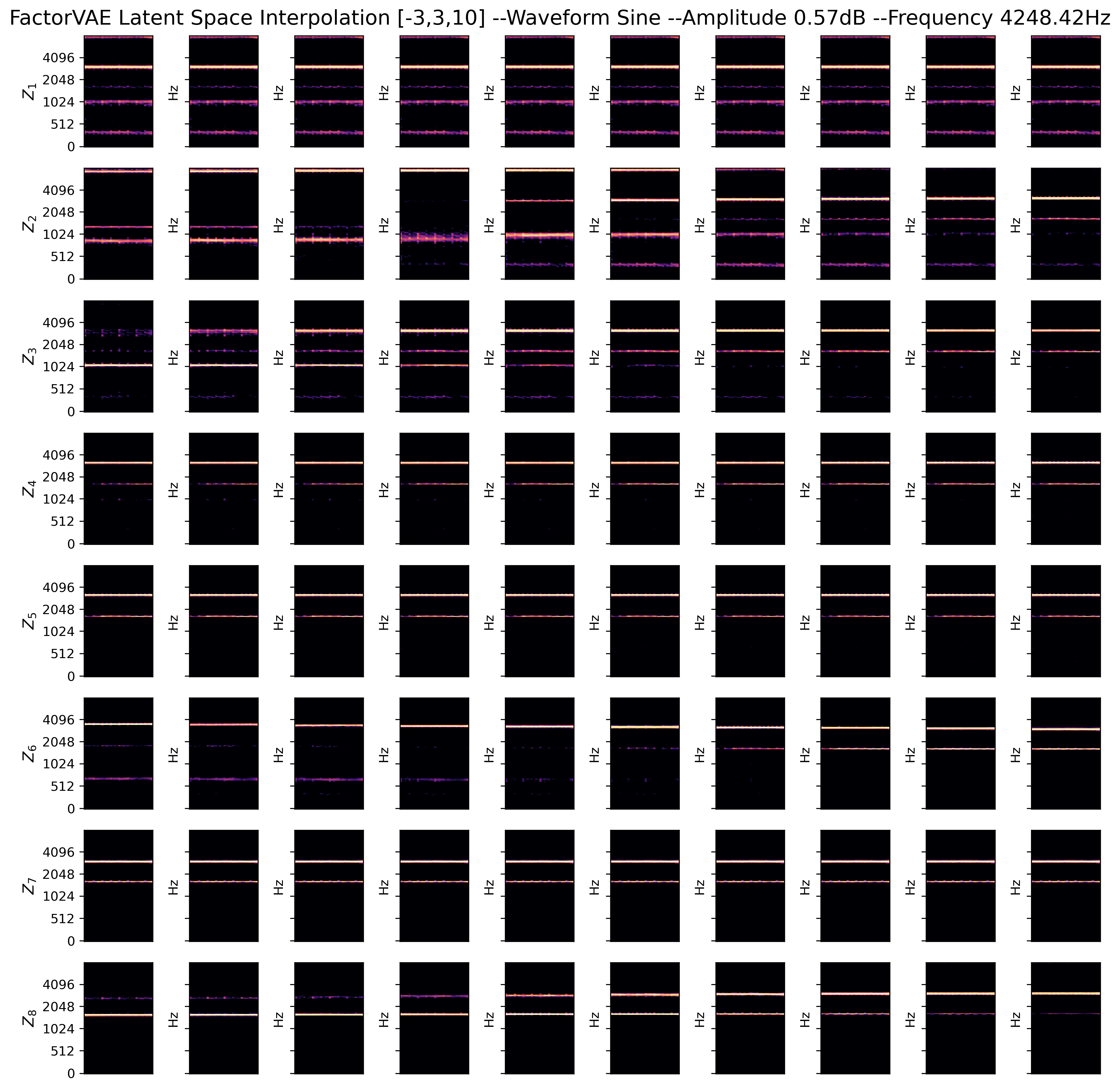}
    \caption{Latent space interpolation for the Factor-VAE model, showcasing variations in a single latent code while others are fixed. Each row represents 10 discrete samples per latent code.}
    \label{fig:FactorVAE_Latent_Space_Interpolation}
\end{figure}

\begin{figure}[!h]
    \centering
    \includegraphics[width=\textwidth]{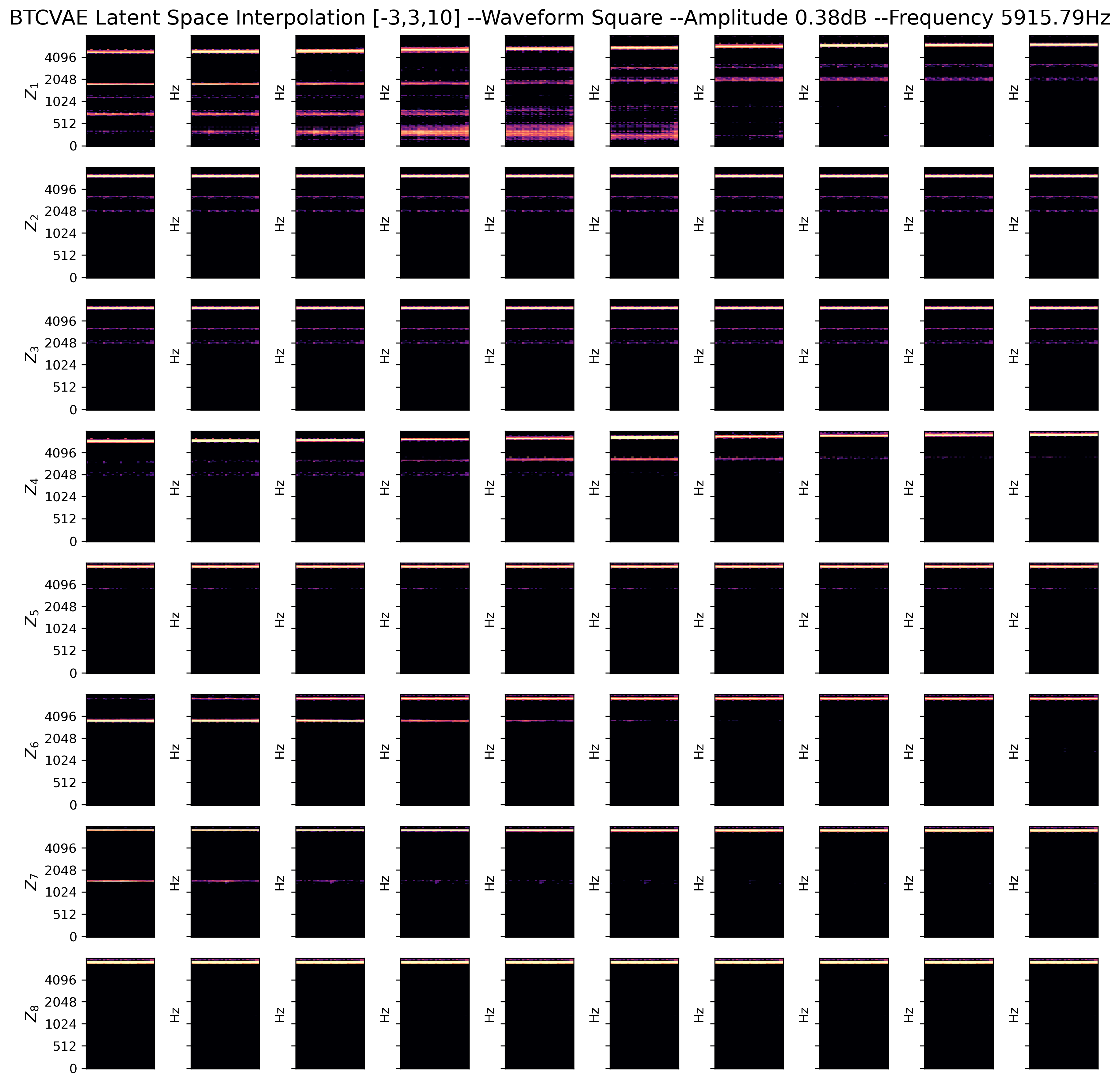}
    \caption{Latent space interpolation for the $\beta$-TCVAE model, showcasing variations in a single latent code while others are fixed. Each row represents 10 discrete samples per latent code.}
    \label{fig:BTCVAE_Latent_Space_Interpolation}
\end{figure}

\end{document}